
\input harvmac
%
%
%
%
%
%
%
%
{\chardef\check0 \immediate\openin\check\jobname.defs
 \ifeof\check\def\readin{\message{\jobname.defs unfound }\relax}%
 \else\def\readin{\message{\jobname.defs found }\input\jobname.defs}\fi
 \expandafter}\readin\relax
\ifx\ymiiba\undefined\message{ymiiba undefined}\def\ymiiba{?}\fi
\writedefs

\def\ymh{Yang-Mills-Higgs}

\def\ib{{\bar I}}
\def\co{{\cal O}}
\def\ie{{\it i.e.}}

\def\cins{constrained instantons}
\def\iib{instanton-antiinstanton pair}

\Title{\vbox{\hbox{UTTG-21-94}\hbox{hep-ph/9502333}}}
     {\centerline{On Constrained-Instanton Valleys}}

\centerline{Meng-yuan Wang\footnote{$^\dagger$}{meng@physics.utexas.edu}}
\bigskip\centerline{Theory Group, Department of Physics,
University of Texas, Austin, TX 78712}

\vskip .3in
We develop a systematic treatment for the quasi-zero modes, which play an
important role in nonabelian gauge theories. It can be used to derive the
analytic forms for the constrained instantons in the \ymh\ theory.
This will automatically sum up contributions to all orders in $\rho$,
the instanton size. We also give the analytic expressions for the
instanton-antiinstanton pair, with an arbitrary relative phase. We
apply the results to the computation for high energy baryon number violating
cross section, as well as to Klinkhamer's new instanton.
\Date{12/94}

\newsec{Introduction}
In the \ymh\ theory where the Higgs mechanism breaks the classical scale
invariance in the gauge sector, instantons are
no longer exact solutions to the field
equations\ref\thooft{G.'t Hooft, {\sl Phys. Rev.} {\bf D14}
(1976) 3432.}. Instead, they represent ``the bottom
of the valley" parametrized by the quasi-zero
mode $\rho$. This valley trajectory
corrresponds to low-lying
configurations which still dominate the path integral.
It is therefore necessary to write down an expression for it if we
are to carry out a reliable semi-classical
approximation for the path integral.
This can be done by introducing some constraints under which
instantons become solutions again. These constrained instantons are first
discussed in a systematic way by
Affleck\ref\affleck{I.Affleck,{\sl Nucl. Phys.} {\bf B191}
(1981) 429.}, who uses an expansion
of the supposedly small parameter $\rho$. Only the leading term,
which is independent of the choice of contraints, is given. Terms of
higher orders in $\rho$ depend on the constraints and are
difficult to evaluate. The author of ref.\xref\affleck\ also has little to say
about the pre-exponential corrections which come with any choice of
constraints. These difficulties can be naturally resolved in the framework
of a new systematic treatment which we shall present in this paper.
In Section 2, we outline this new approach, which is far more powerful
than what was called the valley method previously\nref\yung{A.Yung,
{\sl Nucl. Phys.} {\bf B297} (1988) 47.}\nref\khoze{V.V.Khoze and
A.Ringwald, CERN preprint, CERN-TH-6082-91.}\nref\khozeb{V.V.Khoze and
A.Ringwald, {\sl Nucl. Phys.} {\bf B355} (1991) 351.}\refs{\yung{--}\khozeb},
and derive its
general formulation. In section 3, we apply the results to the \ymh\ system and
give the analytic form for the \cins\ to all orders in $\rho$.

One of the immediate applications of this result is to the process of
high energy anomalous baryon number violation. In order to include
corrections from the final state particles, it is
convenient to employ the optical theorem and consider the
$\langle2\rightarrow2\rangle$
inclusive scattering process. The relevant configuration is therefore
the \iib, which is
another quasi-zero mode  by itself. Previous
works\khozeb\ understandably avoided
dealing directly with
the \ymh\ system, and instead used results in the pure Yang-Mills
theory\yung\ as an approximation. Unfortunately, not only is the contribution
from the Higgs particles poorly estimated, but the gauge sector is also
misrepresented by the wrong valley direction.
We give the analytic expression for
the \ymh\ \iib\ in Section 4. We also clarify some other misconceptions
in the literature on related issues. Multiple \iib{s} can also be
treated in principle. Interested readers
are referred to our other paper\ref\wang{M.Wang, UT preprint UTTG-18-94.}.

It is rather straightforward to write down analytically our
equivalent of Klinkhamer's new instanton\ref\klinkhamer{F.R.Klinkhamer,
{\sl Nucl. Phys.} {\bf B407} (1993) 88.}
using these results. This will be done in Section 5.

\newsec{The Quasi-zero Modes and the Semi-classical Approximation}

Our objective is to evaluate path integrals in Euclidean spacetime using
the semi-classical approximation. For simplicity, all spacetime indices are
suppressed. We will also ignore the need for
gauge fixing, which has been analysed by
Yaffe\ref\yaffe{L.G.Yaffe,{\sl Nucl. Phys.} {\bf B151} (1979) 247.}.
The modification is straightforward and will not be included in this paper.

The semi-classical approximation for the partition function,
\eqn\partition{Z\ =\ \int [{\cal D}A] \exp(-S[A]),}
is given by the Gaussian approximation. One first finds the
minimum of $S[A]$ at $A_0$ by solving the field equation
\eqn\fequation{{{\delta S}\over{\delta A}}\bigg|_{A_0}\ =\ 0,}
 makes a change
of variables to $\delta A
\equiv A-A_0$, and observes that the exponent now becomes
$$-S[A]=
-S[A_0]-{1\over 2}\int d^4 x \delta A {{\delta^2 S}\over{\delta A^2}}\bigg|_
{A_0} \delta A + {\cal O}(\delta A^3).$$
One then proceeds by omitting the higher order terms, which are suppressed by
powers of $g$, the coupling constant, and carrying out the Gaussian integrals.
The end result is,
\eqn\partitiona{Z\ \sim\ {\det}^{-{1\over 2}}\left({{\delta^2 S}\over
{\delta A^2}}\bigg|_{A_0}\right)\exp(-S[A_0]),}
which is the leading term in the expansion with respect to $g$.
The vacuum expectation value of any operator ${\cal X}(A)$ can also be
evaluated in the same manner. We have,
\eqn\vev{\eqalign{\langle{\cal X}\rangle\
&=\ {1 \over Z}\int [{\cal D}A] {\cal X}(A) \exp(-S[A])\cr
&=\ {\cal X}(A_0),\cr}}
again to the leading order in $g$.

Non-trivial solutions to the field equation usually come with zero modes.
They have to be dealt with
using the collective coordinates, which we shall denote as $\omega$.
For example, a Yang-Mills
instanton is parametrized by its size and position. Adding in the
global gauge degrees of freedom, we need eight collective coordinates.
The path integral \partition\ can be evaluated by inserting into it the
identity\foot{Of course for the identity
to hold, we have to assume that for any given
$A$, $<A-A_0^\omega, f_\omega^{(i)}>=0$ has
exactly one solution, $\omega^*(A)$.
However, because we are taking the semi-classical
approximation, this condition in fact
only needs to be imposed on $A$ that is fairly close to $A_0^\omega$.},
\eqn\trick{1\ =\ \int d^n\omega \Delta_\omega \prod_{i=1}^n
\delta(<A-A_0^\omega, f_\omega^{(i)}>),}
where n is the number of collective coordinates, $<X,Y>\equiv
\int d^4x X(x)Y(x)$ is just the ordinary inner product for
functions\foot{We will be using a general constraint, $f$,
thus introducing generalized inner products is unnecessary.}, $f^{(i)}$
are some constraint functions that in general depend on $\omega$.
Notice that $A_0$, being the solution to \fequation, is independent of
the choice of constraints.
The Jacobian $\Delta_\omega$ is given to the leading order in $g$ by,
\eqn\defDelta{\Delta_\omega\ =\ \bigl|\det_{ij}(<{\partial A_0^\omega \over
\partial \omega^{(i)}},f_\omega^{(j)}>)\bigr|.}
The $\delta$-function in \trick\ can be further massaged using
\eqn\mdelta{\delta(x)\ =\ \lim_{\alpha\rightarrow 0}{1\over\sqrt{2\pi\alpha}}
\exp\left(-{1\over{2\alpha}}x^2\right).}
This gives
\eqn\partitionb{Z\ =\ \int d^n\omega Z_\omega,}
and
\eqn\partitionc{\eqalign{Z_\omega\ &\sim\ e^{-S(A_0^\omega)}\Delta_\omega
\lim_{\alpha\rightarrow 0}{1\over{(2\pi\alpha)^{n\over 2}}}
\int [{\cal D}(\delta A)]\exp(-{1\over 2}\delta A\ H\ \delta A)\cr
&\sim\ e^{-S(A_0^\omega)}\Delta_\omega
\lim_{\alpha\rightarrow 0}{1\over{(2\pi\alpha)}^{n\over 2}}
\det\left(H\right)^{-{1\over2}},\cr}}
where
\eqn\defH{H\ =\ {{\delta^2 S}\over{\delta A^2}}
\bigg|_{A_0}+{1\over\alpha}\sum_i
f_\omega^{(i)} \times f_\omega^{(i)}.}
Since the operator ${{\delta^2 S}\over{\delta A^2}}\big|_{A_0}$ has zero modes
${\partial A_0^\omega \over\partial \omega^{(i)}}$, we modify the identity
\eqn\matrixtrick{\det(M+\sum_ib^i\times b^i)\ =\ \det M \det_{ij}(\delta_{ij}
+<b^i, M^{-1} b^j>)}
into
\eqn\mattrick{\det(M+\sum_ib^i\times b^i)\
=\ {\det}^\prime M \det_{ik}<b^i,a^k>
{\det_{kl}}^{-1}<a^k,a^l> \det_{lj}<a^l,b^j>,}
where $M$ is any given matrix, $b$ is a given set of $n$ vectors,
$a$ denotes the $n$ zero modes of $M$ and ${\det}^\prime$ denotes
the nonzero part of the determinant. We finally arrive at
\eqn\partitiond{Z_\omega\ \sim\ {\det_{ij}}^{1\over 2}
(<{\partial A_0^\omega \over\partial \omega^{(i)}},
{\partial A_0^\omega \over\partial \omega^{(j)}}>)
\ {{\det}^\prime}^{-{1\over 2}}\left({{\delta^2 S}\over{\delta A^2}}\bigg|_
{A_0}\right)e^{-S(A_0^\omega)}.}
This coincides with the result in ref.\xref\yaffe\ except for a minus sign in
their formula, which apparently results
from their not taking the absolute value in
the counterpart of \defDelta. Note that all $f$--dependences cancel among
themselves and as a result, $f$ does not enter the expression at all,
just as one would have expected.

If these zero modes are instead quasi-zero modes
, which will be denoted $\xi$, the same derivation still applies.
Once $A_0$ is written down\foot{To find
$A_0$ is in fact where the real difficulty
lies. We will elaborate on this later.}, we can again
go through \trick\ to \defH.  The only change is that instead of \mattrick,
we use \matrixtrick\ itself. This gives
\eqn\partitione{Z\ =\ \int d^n\xi Z_\xi,}
and
\eqn\partitionf{\eqalign{Z_\xi\ \sim\ &{\det_{ij}}
(<{\partial A_0^\xi \over\partial \xi^{(i)}},f_\xi^{(j)}>)
\ \ \ {\det_{ij}}^{-{1\over 2}}
(<f_\xi^{(i)},\left({{\delta^2 S}\over
{\delta A^2}}\bigg|_{A_0}\right)^{-1}f_\xi^{(j)}>)\cr
&{\det}^{-{1\over 2}}\left({{\delta^2 S}\over{\delta A^2}}\bigg|_
{A_0}\right)\ e^{-S(A_0^\xi)},\cr}}
which differs slightly from the result in ref.\xref\yaffe,
\eqn\partitiong{\eqalign{Z_\xi\ \sim\ &{\det_{ij}}
(-<f_\xi^{(i)},f_\xi^{(j)}>)
\ \ \ {\det_{ij}}^{-{1\over 2}}
(-<f_\xi^{(i)},\left({{\delta^2 S}\over
{\delta A^2}}\bigg|_{A_0}\right)^{-1}f_\xi^{(j)}>)\cr
&{\det}^{-{1\over 2}}\left({{\delta^2 S}\over{\delta A^2}}\bigg|_
{A_0}\right)\ e^{-S(A_0^\xi)}.\cr}}

It is worth pointing out that \partitionf\ should be the exact and whole
leading
term in the semi-classical expansion. Therefore, it cannot depend on
the arbitrary choice of contraints, which apparently determines the
exact form for $A_0^\xi$ and in turn $S(A_0^\xi)$. This implicit
$f$-dependence in the
exponent has to be cancelled by the dependence in the pre-exponential
factor. This fact can be used as a consistency check on \partitionf.
Notice that $A_0^\xi$ should remain unchanged while we make the
substitution $f_\xi(x)\rightarrow c(\xi) f_\xi(x)$, where $c(\xi)$ is
a constant in spacetime. Under this ``transformation", \partitionf\
is invariant, while \partitiong\ obviously is not.

Some readers may be alarmed by the fact that our constraint function $f$
has a dependence on $\xi$, and thus does not divide the entire functional
space into well-defined slices. By this we mean that for any given field
configuration $A$, we should find one unique background field $A_0^\xi$
by demanding $<(A-A_0^\xi),f_\xi>=0$, such that we can consider all the
fields $A$ corresponding to the same $A_0^\xi$ as its quantum fluctuation.
It is such a consideration that
prompts the author of ref.\xref\affleck\ to propose
global constraints (such as $\int \tr F^3\propto\rho^{-2}$).
Unfortunately, they are nonlinear and make solving for $A_0^\xi$ practically
impossible. We circumvent this dilemma by noting that in our semi-classical
method, all configurations far away from the valley trajectory
$A_0^\xi$ are approximated by the Gaussian form anyway. Thus we only
need well-defined slices along a small neighborhood of $A_0^\xi$,
justifying our use of $f_\xi$.

This is not the whole story yet, because
although \partitionf\ is an elegant expression, it is very difficult
to evaluate the pre-exponential factor in practice. One thus uses the
exponential part as an approximation. Ideally one would like to choose
$A_0^\xi$ (or equivalently $f_\xi$) so that the pre-exponential factor
in \partitionf\ is independent of $\xi$, but this again is a hopeless
task. The authors of refs. \xref\yung\ and
\xref\khoze\ pursued a ``natural constraint",
\eqn\nconstraint{f_\xi\propto{{\partial A_0^\xi}\over{\partial \xi}},}
which would simplify the expression
a little\foot{The effects of equation \nconstraint\
will be evaluated at the
end of Section 3.}, but does not necessarily minimize the contribution from
the pre-exponential
factor. They did venture out to define a ``generalized" inner product, but
only after using a spherical ansatz to reduce the Yang-Mills lagrangian
to that of a simple
quantum mechanical system for which one has much better physical intuition.
The result still turns out to be undesirable, underscoring the importance of
following our systematic approach. We will elaborate on this
fact in Section 4.
Readers looking for greater details will find them in ref.\xref\wang.

It turns out that our best course of action is to make the best use
of this freedom in choosing the constraint. This would require that we isolate
the $f$-dependent part of $A_0^\xi$, make a reasonable guess for it, add back
in the $f$-independent part, and then find the corresponding constraint $f$.
In practice, we first have to identify the quasi-zero modes that
need to be treated. Next, we write down the set of criteria for
desirable solutions $A_0^\xi$. We then try to find a solution which satisfies
all the criteria. These tasks are far from trivial, but still much more
manageable than to solve $A_0^\xi$ for a given $f_\xi$.
To find the corresponding constraint function $f_\xi$ for
a given $A_0^\xi$ is relatively easy because basically we are looking
for such an $f_\xi$ that
\eqn\newfequation{<{{\delta S}\over{\delta A}}
\bigg|_{A_0^\xi}, \delta A>\ =\ 0,
\hbox{\ for any\ }\delta A\hbox{\ \ satisfying\ }<f_\xi, \delta A>=0.}
Apparently we should
always choose one of the $f$'s to be ${{\delta S}\over{\delta A}}\big|_
{A_0^\xi}$. The choice for the rest of the $f$'s is again arbitrary and
does not affect the final outcome. Notice that this represents a dramatic
change from the way we treat the field equation \fequation.
We need to solve \fequation\ for $A_0$. We can also do the same with
\newfequation\ if given $f$, but the alternative way is to use it to
identify the corresponding $f$
for any given $A_0$. As explained above, this is much easier.

We also want to reemphasize that unlike in the case of zero modes, there is no
such thing as an ``exact solution" for quasi-zero modes a priori. All
trajectories will yield the same correct answer in \partitionf. Only when we
ignore part of \partitionf\ do we have a preference for those which
minimize the effects of the ignored part. Even then we still cannot
say which solution is best exactly because of the uncertainty introduced
by discarding the incalculable terms. But for problems that require only a
qualitative or partially quantitative understanding, the valley method remains
a useful tool. We demonstrate our systematic
approach for the \ymh\ system in the following sections. Detailed
discussions for the pure Yang-Mills theory can be found in ref.\xref\wang.

\newsec{The Constrained Instanton in the \ymh\ Theory}

As mentioned earlier, in the instanton number $Q=1$ sector
the lowest lying configuration of the pure
Yang-Mills action is an instanton with
$8$ zero-modes. If we add Higgs and its
vacuum expectation value into the system,
the scale invariance
is broken, and the $\rho$ direction is lifted to a quasi-zero
mode. This can be seen by observing the effect the rescaling
$A(x)\rightarrow aA(ax)$, $\phi(x)\rightarrow\phi(ax)$ has on the
action\refs{\thooft,\affleck}.
The constrained instanton in the \ymh\ theory is thus
obviously a valley trajectory parametrized by $\rho$.

After identifying the valley direction, our next task is to set up criteria
for the desirable solution $\{A_0^\rho,\phi_0^\rho\}$. Guided by our
earlier discussion and previous experience with the \iib\ in the
pure Yang-Mills theory\wang, we find the following,
\vskip 0.2in
\item{1)} $\{A_0^\rho,\phi_0^\rho\}$ belongs in the $Q=1$ sector.
\item{2)} $\{A_0^\rho,\phi_0^\rho\}$ has easily identifiable instanton
parameters, and covers the entire 8-dimensional
parameter space spanned by these zero- and nonzero-modes.
\item{3)} $\{A_0^\rho,\phi_0^\rho\}$ conforms to all
constraint-independent results, which can usually be obtained by taking
the parameters to certain limits.
\item{4)} $\{A_0^\rho,\phi_0^\rho\}$ respects the symmetries
of the theory.
\vskip 0.2in
\noindent
Just like in the pure Yang-Mills $I\bar I$ case, $Cri.3$ is the most
restrictive and therefore the most useful for our purpose. There is a subtlety
here however. The Higgs vacuum expectation value
$\langle\phi\rangle$, which explicitly breaks the $\rho$ invariance in the
action, should have an interesting interplay with $\rho$ when we take the
limits. We now examine this in detail.

Consider the rescaled \ymh\ action
\eqn\action{S\ =\ {1\over g^2}\int d^4x \left\{ {1\over2}{tr}F^2
+\kappa\left[\big|D\phi\big|^2+{1\over4}\left(\big|\phi\big|^2-
\langle\phi\rangle^2\right)^2\right]\right\},}
where $\kappa$ is the ratio between gauge and Higgs couplings.
We then introduce the following ansatz\foot{The use of this ansatz is by no
means essential to our reasoning, but will serve to simplify the
presentation greatly.}
\eqn\ansatz{A_\mu(x)\ =\ {{x^\nu\Lambda_{\mu\nu}}\over x^2}B(r),\hbox{\ \ \ \ }
\phi(x)\ =\ (1-H(r))\langle\phi\rangle,}
where $r=\big|x\big|$, $\Lambda_{\mu\nu}={1\over{4i}}(\sigma_\mu\bar\sigma_\nu
-\sigma_\nu\bar\sigma_\mu)$\ and $\sigma_\mu=({\vec \sigma},i)$,
$\bar\sigma_\mu=({\vec \sigma},-i)$. The only dimensional variables in
the theory are $r$, $\rho$, $\langle\phi\rangle$
and the renormalization scale $\mu$.
Knowing that effects of $\mu$ can always be absorbed into the renormalization
of the coupling constant, the dimensionless $B$\ and $H$ can only depend on
the variables, $\rho\over r$ and $\langle\phi\rangle r$.
Imagine the double expansions of $B$ and $H$ with respect to these two
parameters\foot{If we require that at
least for some values of $r$, both $\rho\over r$ and $\langle\phi\rangle r$
are small, we will have to assume
that $\rho\langle\phi\rangle$ is small, but we can always push our result
to the $\rho\langle\phi\rangle\ge1$ region later. For
$\rho\langle\phi\rangle\gg1$, the renormalized coupling constant becomes
large, invalidating
our semi-classical approximation, but contributions from that part of phase
space should be highly suppressed anyway.}
\eqn\dexpansion{\eqalign{B({\rho\over r},\langle\phi\rangle r)\ &=\
B_0({\rho\over r})+(\langle\phi\rangle r)^2B_1({\rho\over r})+\cdots\cr
&=\ ({\rho\over r})^2B^0(\langle\phi\rangle r)+({\rho\over r})^4
B^1(\langle\phi\rangle r)+\cdots,\cr
H({\rho\over r},\langle\phi\rangle r)\ &=\
H_0({\rho\over r})+(\langle\phi\rangle r)^2\ln\langle\phi\rangle r
H_1({\rho\over r})+\cdots\cr
&=\ ({\rho\over r})^2 H^0(\langle\phi\rangle r)+({\rho\over r})^4
H^1(\langle\phi\rangle r)+\cdots,\cr}}
in which $B_i$ and $H_i$ will be considered ``row vectors", and $B^j$, $H^j$
``column vectors". Notice that these expansions are not analytic. They contain
logarithms. The exact forms of the expansions can be inferred from expanding
$B_0$, $H_0$ with respect to $\rho\over r$, and $B^0$, $H^0$ with respect to
$\langle\phi\rangle r$, which are exactly calculable.
We now compute the leading row by taking the $r$ (or equivalently
$\langle\phi\rangle$) $\rightarrow0$ limit and
the leading column by taking $r\rightarrow\infty$. We have the
following constraint-independent results\nref\espinosa{O.Espinosa,
{\sl Nucl. Phys.} {\bf B343} (1990) 310.}\refs{\affleck,\espinosa}
\eqn\frow{\eqalign{&B_0\ =\ {{2\rho^2}\over{r^2+\rho^2}},\cr
&H_0\ =\ 1-\left({r^2 \over {r^2+\rho^2}}\right)^{1\over 2},\cr}}
and
\eqn\fcolumn{\eqalign{&B^0\ =\ m_W^2r^2K_2(m_Wr),\cr
&H^0\ =\ {{m_Hr}\over 2}K_1(m_Hr),\cr}}
where $m_W=\sqrt{\kappa\over2}
\langle\phi\rangle$ and $m_H=\langle\phi\rangle$ are the gauge boson and Higgs
masses respectively, and $K_\nu$ is
the modified Bessel function of the second kind,
with the following small $y$ behaviors,
\eqn\sbessel{\eqalign{K_1(y)\
&\sim\ {1\over y} + {\cal O}\left(y\ln{y}\right)\cr
K_2(y)\ &\sim\ {2\over{y^2}} -{1\over2}+{\cal O}\left(y^2\ln{y}\right).\cr}}

It was shown in
ref.\xref\affleck\ that all higher order terms
have dependence on the constraint we choose. Since by our valley method, all
constraint-dependent terms can be freely chosen first and used to determine the
corresponding constraint later, we
can now fill in the blank in any way we like.
How does one fill in the blank in a matrix $M$ when the first row $M_0$ and
the first column $M^0$ are given? Well, the simplest choice seems to be
\eqn\mgeneration{M\ \equiv\ {1\over M_0^0}M^0\times M_0,}
which corresponds to the following choice
\eqn\ymhcia{\eqalign{&B(r)\ =\ {{r^2\rho^2m_W^2}\over{r^2+\rho^2}}K_2(m_Wr),\cr
&H(r)\ =\ \left[1-\left({r^2 \over {r^2+\rho^2}}\right)^{1\over 2}\right]
m_HrK_1(m_Hr),\cr}}
or
\eqn\ymhci{\eqalign{&A_\mu(x)\ =\ {{x_\nu\Lambda_{\mu\nu}\rho^2m_W^2}\over
{x^2+\rho^2}}K_2(m_W\big|x\big|),\cr
&\phi(x)\ =\ \left\{1-\left[1-\left({x^2 \over
{x^2+\rho^2}}\right)^{1\over 2}\right]
m_H\big|x\big|K_1(m_H\big|x\big|)\right\}\langle\phi\rangle.\cr}}

Of course, there are actually infinitely many ways to form a matrix with
only one row and one column fixed. For any background matrix $M_B$, define
$\Delta M^0=M^0-M_B^0$ and $\Delta M_0=M_0-{M_B}_0$. It is obvious that
the choice
\eqn\bgmg{M_B+{1\over {\Delta M_0^0}}\Delta M^0\times \Delta M_0}
would work just as well\foot{In fact,
the converse is also true. For any matrix $M$,
we can always find a certain (but not unique) $M_B$ such that the expression
\bgmg\ equals $M$.}. This would seem to
indicate that there is nothing special about
\ymhci. For example, we may just as well
set all constraint-dependent terms to zero, which corresponds to
\eqn\ymhcib{\eqalign{&A_\mu(x)\ =\ {{x_\nu\Lambda_{\mu\nu}}\over x^2}
\left[\rho^2m_W^2K_2(m_W\big|x\big|)-
{{2\rho^4}\over{x^2(x^2+\rho^2)}}\right],\cr
&\phi(x)\ =\ \left[\left({x^2 \over {x^2+\rho^2}}\right)^{1\over 2}
+{\rho^2\over{2x^2}}-m_Hr\big|x\big|K_1(m_H\big|x\big|)\right]
\langle\phi\rangle.\cr}}
Strangely \ymhcib\ has the wrong asyptotic
behaviors, and cannot be considered a good solution. We seem to have
a paradox at hand. Of course the cause of this paradox is that we have
lied. We are not really free to choose
all the higher order terms in \dexpansion\
even though the dependence on the constraint is genuine. This is because
there is an implicit requirement that these higher order terms cannot dominate
over the preceding terms. To put it more precisely, the ratio ${B_{i+1}\over
B_i}$ better be bounded above for all values of $r$. The same goes for
${B^{i+1}\over B^i}$, ${H_{i+1}\over H_i}$ and ${H^{i+1}\over H^i}$.
In effect, \ymhci\ corresponds to setting all these ratios to constants.
Now it is no longer trivial to write down alternatives to \ymhci, although
they certainly exist in principle. In fact, we will now introduce one of these
alternative expressions which is more
elegant, especially when used to construct
the \iib\ solution in the next section.
\eqn\ymhcix{\eqalign{&A_\mu(x)\ =\
{{x_\nu\Lambda_{\mu\nu}\rho^2m_W^2K_2(m_W\big|x\big|)}
\over{x^2+\rho^2m_W^2x^2K_2(m_W\big|x\big|)/2}},\cr
&\phi(x)\ =\ \left({x^2 \over {x^2+\rho^2m_H\big|x\big|K_1(m_H\big|x\big|)}}
\right)^{1\over 2}\langle\phi\rangle.\cr}}
This is what we consider the best
valley solution for a constrained instanton. One can
easily check that all criteria are satisfied.

Now that we have found the valley configuration, we will proceed to
find the corresponding constraint. As explained in the previous section,
this is rather straightforward, especially since there is only one
quasi-zero mode. $f_\rho$ is none other than ${{\delta S}\over{\delta A}}\big|_
{A_0^\rho}$, where $A$ stands for both $A_\mu$ and $\phi$, and $A_0^\rho$
is \ymhci. Substituting this into
\partitionf, we have
\eqn\partitionh{\eqalign{Z\ =\ \int d\rho &\ \ Z_\rho\cr
\sim\ \int d\rho &\ <{{\partial A_0^\rho} \over{\partial \rho}},
{{\delta S}\over{\delta A}}\bigg|_{A_0^\rho}>\ \
<{{\delta S}\over{\delta A}}\bigg|_{A_0^\rho},\left({{\delta^2 S}\over
{\delta A^2}}\bigg|_{A_0}\right)^{-1}{{\delta S}\over{\delta A}}\bigg|_
{A_0^\rho}>^{-{1\over 2}}\cr
&\ \ {\det}^{-{1\over 2}}\left({{\delta^2 S}\over{\delta A^2}}\bigg|_
{A_0}\right)\ e^{-S(A_0^\rho)}\cr
=\ \int d\rho &\ \ {{dS}\over{d\rho}}\ \
\left|{{\delta S}\over{\delta A}}\bigg|_{A_0^\rho}\right|^{-1}
{{\det}^{\prime\prime}}^{-{1\over 2}}
\left({{\delta^2 S}\over{\delta A^2}}\bigg|_
{A_0}\right)\ e^{-S(A_0^\rho)},\cr}}
where ${\det}^{\prime\prime}$ is the determinant taken in the slices defined
by the constraint\foot{Here we make a comparison with the nonlinear constraint
$\delta\left(\int d^4x\left[{\cal O}(A)-{\cal O}(\rho)\right]\right)$
used in ref.\xref\affleck. To leading order in $g$, this reduces to $\delta(<
{{\delta {\cal O}}\over {\delta A}}\big|_{A_0},\delta A>)$. Comparing this
with \newfequation, we see that our constraint corresponds to choosing $\cal O$
to be the lagrangian. Unfortunately, this implies that all field configurations
in a slice defined by this constraint have the same action.
Therefore we shouldn't be able to solve for
$A_0$, which is by definition the configuration with the lowest action
in the slice, unless we switch back
to our linear version of the constraint. Naively,
if we ignore this fact and go through the formal maneuver of algebra, we
find, for example, the equation for $B_1$ is
$${{\delta^2 S}\over{\delta A^2}}
\bigg|_{B_0,H_0}\left(B_1-\sigma_1 B_0\right)\ =\ 0,$$
where $\sigma_1$ is the leading term in the lagrangian multiplier.
Since our solution \ymhcia\ corresponds to
$$B_1\ \propto\ B_0,$$
it is satisfactory.}.

It would also be interesting to see
what we could get by imposing the ``natural
constraint"
condition \nconstraint. This would
mean abandoning \ymhci\ and instead pursuing the
solution $A_0$ which satisfies
\eqn\nconstrainta{{{\delta S}\over{\delta A}}\bigg|_{A_0^\rho}
\propto{{\partial A_0^\rho}\over{\partial \rho}}.}
Let's suppose for a moment that such a solution could indeed be found.
Equation \partitionh\ will then be slightly simplified. We have
\eqn\partitioni{Z\ \sim\ \int d\rho
<{\partial A_0^\rho \over\partial \rho},
{\partial A_0^\rho \over\partial \rho}>^{1\over 2}
\ {{\det}^{\prime\prime}}^{-{1\over 2}}
\left({{\delta^2 S}\over{\delta A^2}}\bigg|_
{A_0}\right)e^{-S(A_0^\rho)},}
which is very similar to \partitiond, our formula for exact zero modes. The
first term in the integral can be easily calculated if we are given
$A_0^\rho$. However, the second term,
which is probably beyond our reach, also has a nontrivial dependence on
$\rho$ which is not necessarily negligible.
Since in reality we cannot even solve for
$A_0^\rho$ under \nconstrainta, we have nothing
to gain by taking this approach.

\newsec{Baryon Number Violation and the Instanton-Antiinstanton Pair}
It was first suggested by Ringwald\ref\ringwald{A.Ringwald,
{\sl Nucl. Phys.} {\bf B330}
(1990) 1.}\ and later Espinosa \espinosa\ that anomalous
baryon number violation in the
standard model might be observable in high-energy colliders.
They, and other authors later,
suggested that at nonzero
energy, the 't Hooft suppression factor\thooft\
$\exp(-4\pi/\alpha)$ for this process
is replaced by
\eqn\hg{\sigma\ \sim\ \exp\left\{\left({4\pi\over\alpha}\right)
F_{hg}\left(\left({E\over E_{sphal}}
\right)^{2\over3}\right)\right\},}
where $E_{sphal}\sim m_W/\alpha$ is the
sphaleron scale, and $F_{hg}$ is the so-called
``holy grail" function which increases
from $-1$ at zero energy to values presumably
closer to $0$ near $E_{sphal}$.
These pioneering works are
based on single-instanton calculations
and do not take into account corrections from
the incoming particles, the outgoing particles and multiple instanton
effects\nref\mattis{M.Mattis,
{\sl Phys. Rep.} {\bf 214} (1992) 159. This is an excellent
review which much of our understanding on recent state of research in
baryon number violation is
based on.}\nref\guida{R.Guida, K.Konishi and N.Magnoli,
{\sl Int. J. Mod. Phys.}
{\bf A9} (1994) 795.}\refs{\mattis,\guida}. We have
very little to say about the incoming particle
corrections\nref\hsu{S.Hsu, Preprint
HEP-PH/9406234.}\nref\mclerran{M.Mattis, L.McLerran and L.Yaffe,
{\sl Phys. Rev.}
{\bf D45} (1992) 4294.}\refs{\hsu,\mclerran}, and will
concentrate our attention on the contributions
coming from the \iib\nref\zakharov{V.Zakharov, TPI
preprint TPI-MINN-90/7-T.}\nref\porrati{M.Porrati, {\sl
Nucl. Phys.} {\bf B347} (1990) 371.}\refs{\khozeb,\zakharov,\porrati},
which has
been shown to sum up the outgoing particle corrections perturbatively to all
orders of $E\over E_{sphal}$\ref\arnold{P.Arnold and M.Mattis,
{\sl Phys. Rev.}
{\bf D44} (1991) 3650.}. Little progress has
been made so far on multiple instanton
effects due to the difficulty of finding expressions corresponding to multiple
instantons
and antiinstantons. Our earlier paper\wang\ gives
these expressions for the pure
Yang-Mills theory. It would be interesting
to carry over the results to the \ymh\ theory,
but the subject is really beyond the scope of this paper and will have
to await future efforts.

The idea that puts the valley method and
the baryon number violation together is
the following. Instead of calculating
the $\langle2\rightarrow$any$\rangle$ cross
section one by
one, one applies the optical theorem and
computes the imaginary part of the forward
$\langle2\rightarrow2\rangle$ amplitude. Since we are
only interested in the anomalous part of the
total amplitude, the idea is to consider the $\langle2\rightarrow2\rangle$
amplitude in the
presence of an \iib\ background. This leap of faith turns out to be valid,
at least in the sense that it provides
the analytic continuation of $F_{hg}$ from its
low-energy limit\arnold. The actual computation is still difficult. Since so
little was known about the \ymh\ $I\bar I$ configuration at the time,
these authors\khozeb\ resorted to using the action $S_{\rm Yung}$
of an available pure Yang-Mills valley,
together with the leading contribution from
the Higgs sectors of a single
constrained \ymh\ instanton and an antiinstanton. They
therefore had
\eqn\krmodel{\sigma\ \sim\ {\rm Im}
\int dR\ d\rho_1\ d\rho_2\ \exp[ER-\pi^2(\rho_1^2
+\rho_2^2)\langle\phi\rangle^2-S_{\rm Yung}],}
where $R$ is the separation between $I$
and $\bar I$, and $\rho_1$ and $\rho_2$ are
the sizes of the instanton and the antiinstanton respectively. The first term
in the exponent is the naive initial state phase factor. After the change of
 variables,
\eqn\chvariable{\eqalign{&\theta=R/\rho,\ \ \ \ \ \ \ \ \ \ \ \ \ \ \ \
\bar\rho_i={1\over2}g\langle\phi\rangle\rho,\cr
&\bar E=gE/8\pi^2\langle\phi\rangle,\ \ \ \ \ \
\bar S=(g^2/16\pi^2)S_{\rm Yung},\cr}}
they proceeded with a saddle-point analysis. Assuming $\rho_1=\rho_2\equiv\rho$
at the saddle point, they arrived at
\eqn\sp{\bar\rho=\bar E\theta,\ \ \ \bar S^\prime(\theta)=\bar E\bar\rho.}
Eliminating $\bar\rho$ then leaves
\eqn\spa{\bar E^2\theta\ =\ \bar S^\prime(\theta).}
Because $S_{\rm Yung}$ is exactly known, \spa\ can be solved either
graphically or numerically.
For small $E$, the large $\theta$ expansion of $\bar S$,
\eqn\sexpansion{\bar S(\theta)\ \rightarrow\ 1\ -\ 6/\theta^4\
+\ O(\theta^{-6}),}
can be used to solve for $\theta(\bar E)$. Substituting the result
back into \krmodel, they had
\eqn\hga{F_{hg}\ \sim\ -1+{9\over8}
\left({E\over E_{sphal}}\right)^{4\over3}+\cdots,}
in agreement with the results previously derived using other methods.

As energy increases towards a finite critical
energy $E_{crit}$, which is determined
by the slope of $\bar S^\prime$ at $\theta=0$,
the solution $\theta(\bar E)$ tends
to zero, hence $F_{hg}$ approaches zero as well.
This seems to indicate that $E_{crit}$
is the energy at which anomalous processes are
no longer suppressed. Simple algebra
shows that for small $\theta$'s,
\eqn\sexpansiona{\bar S(\theta)\ =\ {6\over5}\theta^2-{4\over5}\theta^3
+{9\over{35}}\theta^4+\cdots.}
$E_{crit}$ is determined by the coefficient of the first term,
\eqn\ecritical{E_{crit}\ =\ \sqrt{12\over5}E_{sphal}.}

It is probably not surprising that this analysis
was greeted with some skepticism.
Some authors argued that for small values of $\theta$, the
valley configuration should be considered a quantum fluctuation of the
trivial vacuum. Therefore it contributes to the
non-anomalous cross section instead
of the anomalous one. This is a valid concern, and
also a source of great confusion
in the literature. (There have been some papers which argue that cuts or poles
exist to separate the anomalous part from the non-anomalous one in the valley
approach.) Our view on this issue is the following.
There are three relevant versions of
the optical theorem. The general version is
all-inclusive, and thus mixes the anomalous
part with the non-anomalous one, and the multi-instanton contributions with the
single-instanton ones. Obviously it is not useful for our purpose.
There is also a perturbative optical theorem based on Cutkosky's cutting
rules\nref\cutkosky{R.Cutkosky, {\sl J. Math. Phys.} {\bf 1}
(1960) 429.}\nref\veltman{G.'t Hooft and T.Veltman,
``Diagrammar", in: ``particle
Interactions at Very High Energies, Part B", eds.
D.Speiser et al.(1974)pp.210.}\refs{\cutkosky,\veltman}.
It was originally applied
to the trivial-vacuum background, and simply stated that
the optical theorem still
holds when restricted to perturbative non-anomalous cross-section.
As far as the valley method is concerned, this corresponds to expanding
$\bar S(\theta)$ in $\theta$ at the origin.
Therefore eq.\sexpansiona\ just gives
the non-anomalous result in ordinary perturbation theory.
The third version of the optical theorem was developed specifically for
the anomalous processes\arnold. These authors applied the cutting rules to
a widely-separated $I\bar I$ background configuration. They showed that
the anomalous cross-section can also be related to the imaginary part
of the $\langle2\rightarrow2\rangle$ amplitude in a perturbation series.
The expansion parameter is $1/\theta$, and the expansion point is at
$1/\theta=0$(see \sexpansion). In terms of
$F_{hg}$, this is equivalent to an expansion
in powers of $E$. Clearly, both the non-anomalous
and the anomalous cross-sections
calculated this way are perturbative in nature, and are valid only within the
range of validity of the expansion
series, \sexpansiona\ and \sexpansion\ respectively.
Since the two do not intersect each other, there is no paradox.

Before we proceed to derive an expression
for the \iib, which in principle will allow
us to compute the anomalous cross-section
perturbatively to all orders in $E$, let's
first assume that for some unknown reason,
the analysis leading to eq.\ecritical\
has its merits and is worth refining.
In fact, several authors have made the extra effort to criticize
\ecritical\ on technical grounds. Unfortunately these criticisms themselves are
often prone to errors. For example, the
author of ref.\xref\mattis\ argued that
the freedom in choosing the constraint allows one to find valleys which give
different $\bar S(\theta)$'s. Since $E_{crit}$
is determined by $\bar S^{\prime\prime}$,
we won't have a definite prediction for $E_{crit}$. To make matters worse,
the analysis leading to \ecritical\ assumes
that $\bar S^{\prime\prime\prime} < 0$.
When $\bar S^{\prime\prime\prime}$ is made
positive, the solution ``bifurcates".
These problems were then incorrectly attributed to
the unknown incoming particle contributions.

Based on our previous discussion, we know that these inconsistencies should be
resolved within the scope of \partitionf.
A valley is called a valley because it rises slower than quadratic.
Therefore if one encounters positive
third derivatives, it is obvious that the valley
trajectory has been so badly chosen that previously discarded pre-exponential
factors now dominate. In fact, the third derivative of $\bar S$ should always
be zero. It is the fourth derivative
that will remain negative. This is the result
of a $Z_2$ symmetry\foot{Remember that $\theta$ corresponds to the
quasi-zero mode $\big|\vec R\big|$,
where $\vec R$ is the displacement between $I$
and $\bar I$. $\vec R$ has four degrees
of freedom, the other three being zero-modes
of the theory. If we require $S$ to be a
smooth function of $\vec R$, in particular
at the origin, we find that we need to impose the $Z_2$ symmetry.}
, by which I mean $\bar S$ should be invariant
under $\theta\rightarrow-\theta$.
Therefore Yung's $I\bar I$ solution does not even qualify
for a good valley for the pure Yang-Mills
theory (Remember $Cri.3$?). Fortunately
we have found another better solution
$A_{I\bar I}^{\rm YM}$ which we shall derive
later in \ymiiba. It gives
\eqn\sexpansionb{\bar S(\theta)\ =\
{6\over5}\theta^2-{{33}\over{35}}\theta^4+\cdots.}
Notice that the transformation used
in ref.\xref\mattis\ to demonstrate the possibility of
changing $\bar S^{\prime\prime}$, conflicts with the $Z_2$ symmetry mentioned
above. Therefore it cannot be applied on our $A_{I\bar I}$ to generate other
solutions.

We now begin our quest for the expression $A_{I\bar I}^{\rm YMH}$
for the \ymh\ \iib, which should enable us to calculate not only $h$
but also the whole action to all orders in $\rho$.
We will start by reviewing the derivation for $A_{I\bar I}^{\rm YM}$.
Notice that classically the 4-dimensional
pure Yang-Mills theory has the conformal
group as its symmetry. Therefore those 15
generators of the conformal group should
correspond to zero-modes. Unfortunately
the analysis is more complicated than one
would have expected because some of them become gauge zero-modes
sometimes,\ie\ some conformal transformations can be undone by gauge
transformations. For the $I\bar I$ configuration, we start with 16 parameters,
consisting of 8 positions, 2 sizes
and 6 phases. After some tedious examination,
we find that
all of them correspond to zero-modes except for a mixed parameter $z\ =\
(R^2+\rho_1^2+\rho_2^2)/2\rho_1\rho_2$, which can be interpreted as
the separation between $I\bar I$,
and one relative phase, which we will ignore for now.
Just like the constrained
instanton in the \ymh\ theory we discussed in Section 3,
we first write down the four criteria,
\vskip 0.2in
\item{1)} $A_{I\bar I}^{\rm YM}$ belongs in the $Q=0$ sector.
\item{2)} $A_{I\bar I}^{\rm YM}$ has easily identifiable instanton
parameters, and covers the entire 16-dimensional
parameter space spanned by the zero- and nonzero-modes, although we will not
worry about the phases until later.
\item{3)} $A_{I\bar I}^{\rm YM}$ conforms to all
constraint-independent results. In this case, we in fact require that
\itemitem{3.1)} $A_{I\bar I}^{\rm YM}$
becomes the sum of an instanton and an antiinstanton for large $z$.
\itemitem{3.2)} $A_{I\bar I}^{\rm YM}$ vanishes
as $z\rightarrow0$.
\item{4)} $A_{I\bar I}^{\rm YM}$ respects the symmetries
of the theory.
\vskip 0.2in

Obviously the most straightforward way to satisfy $Cri.2$ and $Cri.3.1$
is to write $A_{I\bar I}^{YM}$ as the linear combination of
an instanton and an antiinstanton. Just putting the two objects together
doesn't work well though, as many earlier
researchers discovered to their dismay.
The difficulty lies in the small $z$ behavior. It is harder than it looks to
make such a linear expression vanish, especially if one insists on having the
two objects in the same gauge, which most of these authors did.
An interesting phenomenon which first
appeared in $A_{\rm Yung}$ (although no one
realized its importance then) is that if one of the (anti)instantons is set in
the regular
gauge and the other in the singular
gauge, the sum of the two becomes pure gauge
at $z=0$, \ie
\eqn\coincidence{{{2x^\nu\Lambda_{\mu\nu}}\over{x^2+\rho^2}}+
{{2\rho^2x^\nu\Lambda_{\mu\nu}}\over{x^2(x^2+\rho^2)}}\ =\
{{2x^\nu\Lambda_{\mu\nu}}\over{x^2}},}
This pure gauge configuration can then be transformed into the trivial vacuum
by the gauge transformation
\eqn\gt{A\ \rightarrow\ g_0^{-1}Ag_0+g_0^{-1}dg_0,}
where $g_0(x)={{\sum_\nu x^\nu\sigma_\nu}/{\big|x\big|}}$.
This algebraic coincidence strongly suggests that a linear combination
of an antiinstanton at the origin in the regular gauge and
an instanton at $x=\vec R$ in the singular gauge
(or the other way around) should be considered, \ie
\eqn\ymiib{A_\mu\ =\ {{2x^\nu{\Lambda}_{\mu\nu}}\over{x^2+\rho_1^2}}+
{{2\rho_2^2(x^\nu-R^\nu){\Lambda}_{\mu\nu}}
\over{(x-\vec R)^2[(x-\vec R)^2+\rho_2^2]}},}
where $\vec R$ is a 4-vector.
This expression will automatically satisfy $Cri.1,2$ and $3$.
$Cri.4$ is trickier. It implies that all $\{R, \rho_1,\rho_2\}$ of the
same conformal class (\ie\ those that can related to each others by conformal
transformations) should give the same action. Unfortunately, this is not true
with \ymiib. There are three quasi-zero
modes, $R=\big|\vec R\big|$, $\rho_1$ and $\rho_2$,
while we want only one. The simplest way to conform to the conformal symmetry
is the following. We can choose a one-dimensional slice
in the three dimensional parameter space
spanned by $R$, $\rho_1$ and $\rho_2$, use \ymiib\ on this slice only, and
project the solution on this slice to the rest of the parameter space using
the conformal group. If one chooses the slice to be
\eqn\yungslice{\vec R\ =\ 0,\ \ \ \ \ \ \rho_1\rho_2\ =\ 1,}
one recovers Yung's $I\bar I$ solution. If one chooses instead
\eqn\ourslice{\rho_1\ =\ \rho_2\ =\ 1,}
we have the solution which we shall refer to as $A_{I\bar I}^{\rm YM}$
exclusively from now on,
\eqn\ymiiba{A_\mu\ =\ {{2x^\nu{\Lambda}_{\mu\nu}}\over{x^2+1}}+
{{2(x^\nu-R^\nu){\Lambda}_{\mu\nu}}\over{(x-\vec R)^2[(x-\vec R)^2+1]}}.}
Our choice has a few advantages over
Yung's solution. Besides satisfying the $Z_2$
symmetry\foot{As mentioned earlier,
this is the $Z_2$ symmetry of $R\rightarrow -R$.
The slice \ourslice\ has been assumed.
If we choose to work on the \yungslice\ slice,
there is another $Z_2$ which corresponds
to $\rho_1/\rho_2\rightarrow\rho_2/\rho_1$.
It turns out that these two $Z_2$ symmetries
are in fact conformally equivalent,
\ie\ one implies the other if the solutions
on the two slices are related to each other
by conformal projection.}
\foot{There is another $Z_2$ symmetry one can impose. This is the
space-time reflection symmetry of the lagrangian
density for $I\bar I$ of identical
sizes. Unfortunately neither our solution nor Yung's respect this symmetry.
We are not sure whether this constitutes a serious problem.}
mentioned earlier, it directly corresponds to
the relevant quasi-zero modes in its
\ymh\ counterpart, as we shall show below.

In the \ymh\ system, $R$, $\rho_1$ and $\rho_2$
are all quasi-zero modes. Therefore
we shall look for the analog of
\ymiib\foot{Unlike the $\rho$ valley we dealt with
in the previous section, the $R$ valley has other solutions which
one can write down easily\wang.
This doesn't invalidate our results, however. These alternative
valley trajectories always exist, no matter we can find them or not.}.
Notice that \ymhcix\ is written in the
singular gauge. For an constrained antiinstanton in the regular gauge,
we take the
conjugate ($\Lambda\rightarrow\bar\Lambda$) of \ymhcix\ and apply the gauge
transformation \gt\ on it. We have
\eqn\ymhcai{\eqalign{&A_\mu(x)\ =\ {{2x_\nu\Lambda_{\mu\nu}}\over{x^2+
\rho^2m_W^2x^2K_2(m_W\big|x\big|)/2}},\cr
&\phi(x)\ =\ \left(1+{{\rho^2m_HK_1(m_H\big|x\big|)}\over{\big|x\big|}}
\right)^{-{1\over 2}}
{{x^\nu\sigma_\nu}\over{\big|x\big|}}\langle\phi\rangle.\cr}}
Now clearly we need only to put
eq.\ymhcai\ and eq.\ymhcix\ together for the gauge field,
\eqn\ymhiib{\eqalign{A_\mu(x)\ =\ &{{2x_\nu\Lambda_{\mu\nu}}\over{x^2
+\rho_1^2m_W^2x^2K_2(m_W\big|x\big|)/2}}\cr
&+{{(x_\nu-R_\nu)\Lambda_{\mu\nu}\rho_2^2m_W^2K_2(m_W\big|x-R\big|)}\over
{(x-R)^2+\rho_2^2m_W^2(x-R)^2K_2(m_W\big|x-R\big|)/2}},\cr}}
but what do we do with the Higgs?
By examining the term $\big|D\phi\big|^2$ in the action \action, we find that
$\phi$ has to have the same winding number as $A_\mu$ at infinity, \ie
\eqn\ymhiibh{\lim_{x\rightarrow\infty}\phi(x)\ \rightarrow\
{{x^\nu\sigma_\nu}\over{\big|x\big|}}\langle\phi\rangle.}
We shall choose
\eqn\ymhiibha{\phi(x)\ =\ \left[1+{{R^2}\over{1+R^2}}
\left({{\rho^2m_HK_1(m_H\big|x-R\big|)}\over{\big|x-R\big|}}+
{{\rho^2m_HK_1(m_H\big|x\big|)}\over{\big|x\big|}}\right)\right]^{-{1\over 2}}
{{x^\nu\sigma_\nu}\over{\big|x\big|}}\langle\phi\rangle.}
Notice that $\phi= 0$ at the centers of the (anti)instantons, which are not
where the fermion level crossing occurs as calculated in ref.\xref\guida. This
does not seem to be a problem,
because unlike the anthors of ref.\xref\guida, we
do not believe there to be any definite connection
between fermion level crossing
and a vanishing Higgs VEV.

There is a caveat we have to caution the readers of. The $Z_2$ symmetry that
disqualifies
Yung's solution earlier also haunts our $A_{I\bar I}^{\rm YMH}$ in the
$\rho_1/\rho_2$ direction. Fortunately in most cases we are interested in,
the saddle point lies in the symmetric configurations, \ie\ $\rho_1=\rho_2$.
Nonetheless, anyone who is interested in performing calculations based on
our $A_{I\bar I}^{\rm YMH}$ solution should always be aware of this limitation.

What is the action of a widely separated YMH $I\ib$ pair as described by
eq.\ymhiib\ and eq.\ymhiibha? The leading interaction term should contain
exponential factors: $e^{-M_Wr}$ in the gauge part and $e^{-M_Hr}$ in the
Higgs part. Since $M_W<m_H$, we should concentrate on the gauge interaction,
which can be calculated following the derivation of the leading YM $I\ib$
interaction in ref.\nref\callan{C.G.Callan, R.Dashen and D.J.Gross,
{\sl Phys. Rev.} {\bf D17} (1978) 2717.}\xref\callan. Using the large $M_Wr$
approximation for the modified Bessel function,
\eqn\largebessel{K_2(M_Wr)\ \sim\ \left({{\pi}\over{2M_Wr}}\right)^{1\over2}
e^{-M_Wr},}
we find,
\eqn\gaugeint{S_{\rm gauge}\ \sim\ {{16\pi^2}\over{g^2}}\left[1-
3\rho^4\left({{\pi M_W^3}\over{2r^5}}\right)e^{-M_Wr}\right].}
Replacing $S_{\rm Yung}$ with $S_{\rm gauge}$ in eq.\krmodel\ then
leads to
\eqn\barrier{F_{hg}\ \sim\ -1+\co\left({E\over M}ln{M\over E}\right),}
which is typical of barrier penetrations and coincides with the results in
ref.\nref\banks{T.Banks, G.Farrar, M.Dine, D.Karabali and B.Sakita,
{\sl Nucl. Phys.} {\bf B347} (1990) 581.}\xref\banks, but contradicts the
commonly used expansion of $F_{hg}$ in $(E/M)^{2\over3}$.

Of course, the interaction term in eq.\gaugeint\ is nonperturbative, thus
the perturbative optical theorem does not relate the valley result \barrier\
to the inclusive $\langle2\rightarrow 2\rangle$ cross section. However, we
still have to
explain the apparent success of the original valley calculation by Khoze and
Ringwald\khozeb. The paradox is the following. The R-term
method\nref\khlebnikov{S.Khlebnikov, V.Rubakov and P.Tinyakov,
{\sl Nucl. Phys.} {\bf B350} (1991) 441.}\nref\mueller{A.Mueller,
{\sl Nucl. Phys.} {\bf B364} (1991) 109.}\nref\arnoldb{P.Arnold and M.Mattis,
{\sl Mod. Phys. Lett.} {\bf A6} (1991) 2059.}\nref\diakonov{D.Diakonov and
M.Poliakov, {\sl Nucl. Phys.} {\bf B389} (1993)
109.}\nref\silvestrov{P.Silvestrov, {\sl Phys. Lett.} {\bf B323}
(1994) 25.}\refs{\khlebnikov--\silvestrov} incorporates
the large $r$ behavior (which is exponential) of constrained instantons.
How can the valley method\nref\balitsky{I.Balitsky and A.Sch\"afer,
{\sl Nucl. Phys.} {\bf B404} (1993) 639.}\refs{\khozeb,\balitsky}
reproduce the correct results with the 't Hooft
approximation, which decays as powers of $r$ at large distances? The solution
lies in a more careful examination of what the physically relevant limit is.
By this, we mean that we should not just take the limit of
$E\ll M_{sphal}$ but $M_W\ll E\ll M_{sphal}$ because we are interested in
final states with a large number of particles. The corresponding $I\ib$
separation is therefore in the range of $M_Wr\ll 1$, where the 't Hooft
approximation for (anti-)instantons is valid. The original valley
calculation put together a pair of 't Hooft (anti-)instantons and
calculated the constraint-independent terms in a series expansion of $M_Wr$.
This expansion gives the same result as the exact $I\ib$ solution would
in the region of interest. Therefore our exact $I\ib$ expression should
lead to a closed form for $F_{hg}$ which has the small $E$ expansion
\hga. It is well known\nref\khlebnikovb{S.Khlebnikov and P.Tinyakov,
preprint CERN-TH6146/91(1991).}\refs{\arnold,\khlebnikov} that
constraint dependence starts to manifest itself at order
$(E/E_{sphal})^{(8/3)}$. Our formalism should make it easier to find
the corresponding constraints for any given valley. At order
$(E/E_{sphal})^{(10/3)}$ and higher, the initial state corrections kick
in, and the valley method is incomplete.

\newsec{Klinkhamer's New Instanton}
Two years ago, Klinkhamer suggested the possible existence of a new constrained
``instanton" in the \ymh\ theory\nref\klinkhamera{F.R.Klinkhamer,
Preprint hep-ph/9409336.}\refs{\klinkhamer,\klinkhamera},
and its associated sphaleron\ref\klinkhamerb{F.R.Klinkhamer, {\sl Nucl. Phys.}
{\bf B410} (1993) 343.}.
The proof of existence was done using the indirect argument that involves
the construction of a non-contractible loop under a nonlinear constraint a la
Affleck which serves to prevent the collapse of scale.
Using the ansatz,
\eqn\kansatz{\eqalign{A_\mu\ =\ &{\tilde f}(x)
\left[{\Lambda}_{\mu\nu}(x^\nu-R^\nu)+
{\Lambda}_{\tau\nu}x^\nu{1\over2}{\rm tr}
(\sigma_1\sigma_\tau\sigma_1{\bar \sigma_\mu})\right]\cr
\phi\ =\ &{\tilde h}(x)\langle\phi\rangle,\cr}}
where $\sigma_1$ generates the opposite phases, the new constrained
solution was approximated by numerically
varying ${\tilde f},{\tilde h}$ and $R$
to minimize the action under the condition ${\tilde h}(0)={\tilde h}(R)=0$.

It is quite obvious that this solution corresponds to a constrained \iib\ of
separation $R$ and equal sizes $\rho_1=\rho_2=\rho$.
We therefore should look for a way to include in \ymhiib\ a nonzero relative
phase $a=a^\nu\sigma_\nu$, where $a^\nu$ is a unit 4-vector. We choose to put
this phase on the instanton, which is in the singular gauge, so that \ymhiibh\
and \ymhiibha\ will remain unchanged. We therefore modify \ymhiib\ into
\eqn\ymhiiba{\eqalign{A_\mu(x)\ =\ &{{2x_\nu\Lambda_{\mu\nu}}\over{x^2
+\rho_1^2m_W^2x^2K_2(m_W\big|x\big|)/2}}\cr
&+{{(x_\nu-R_\nu){1\over2}{\rm tr}(\bar a\sigma_\tau a\bar\sigma_\mu)
{\Lambda}_{\tau\nu}\rho_2^2m_W^2K_2(m_W\big|x-R\big|)}\over
{(x-R)^2+\rho_2^2m_W^2(x-R)^2K_2(m_W\big|x-R\big|)/2}},\cr}}
For the problem at hand, we set $a^\nu=(1,0,0,0)$ to
generate the opposite phases
needed.
The next step is to evaluate the action profile with
respect to $R$, with $\rho$
fixed at various values. If for a given $\rho$, the
minimal action $S(R^*(\rho))$
has a value below twice the one-instanton action,
$\{\rho,R^*(\rho)\}$ then gives
a new constrained solution.
Unfortunately, we are not equipped for the numerical
computation required for this
task. We present instead the small-$R$ limit of the action for the Yang-Mills
\iib\ with opposite phases,
\eqn\actiona{S(R,\rho)\ \sim\ 2-{1\over3}\left({R\over\rho}\right)^2+\cdots}

We do not expect  the action profile computed from our
\ymhiiba\ to differ greatly
from the results in ref.\xref\klinkhamer. However, they probably will not be
exactly identical either.
Afterall, two very different constraints are used,
and the resulting approximations do not have to coincide with each other
since the pre-exponential factors are again discarded.

\bigbreak\bigskip\bigskip\centerline{{\bf Acknowledgements}}\nobreak
We thank S.Hsu for explaining his work,
and for helpful discussions on several other subjects.
We are also grateful to J.Distler and J.Feinberg for discussions on
the subject of fermion level crossing. We also thank F.R.Klinkhamer
for his feedback after reading the original manuscript.
This research is supported in part by the Robert A. Welch Foundation and
NSF Grant No. PHY9009850.

\listrefs
\bye